\documentclass[runningheads]{llncs}

\usepackage{graphicx}
\usepackage{booktabs}
\usepackage{adjustbox}
\usepackage{algorithm}  
\usepackage{algpseudocode}  
\usepackage{amsmath}  
\usepackage{longtable}
\usepackage{amsmath,amssymb}
\usepackage{mathrsfs}
\usepackage{pifont}  
\usepackage{multirow}
\usepackage{bbding}
\usepackage[figuresleft]{rotating}

\algnewcommand{\IIf}[1]{\State\algorithmicif\ #1\ \algorithmicthen}
\algnewcommand{\EndIIf}{\unskip\ \algorithmicend\ \algorithmicif}
\algnewcommand\algorithmicswitch{\textbf{switch}}
\algnewcommand\algorithmiccase{\textbf{case}}
\algnewcommand\algorithmicassert{\texttt{assert}}
\algnewcommand\Assert[1]{\State \algorithmicassert(#1)}%

\algdef{SE}[SWITCH]{Switch}{EndSwitch}[1]{\algorithmicswitch\ #1\ \algorithmicdo}{\algorithmicend\ \algorithmicswitch}%
\algdef{SE}[CASE]{Case}{EndCase}[1]{\algorithmiccase\ #1}{\algorithmicend\ \algorithmiccase}%
\algtext*{EndSwitch}%
\algtext*{EndCase}%

\usepackage{algorithm,algpseudocode,float}
\usepackage{lipsum}
\usepackage{makecell}
\newcommand{\tabincell}[2]{\begin{tabular}{@{}#1@{}}#2\end{tabular}}
\algrenewcommand\textproc{}% Used to be \textsc

\makeatletter
\newenvironment{breakablealgorithm}
{% \begin{breakablealgorithm}
	\begin{center}
		\refstepcounter{algorithm}% New algorithm
		\hrule height.8pt depth0pt \kern2pt% \@fs@pre for \@fs@ruled
		\renewcommand{\caption}[2][\relax]{% Make a new \caption
			{\raggedright\textbf{\ALG@name~\thealgorithm} ##2\par}%
			\ifx\relax##1\relax % #1 is \relax
			\addcontentsline{loa}{algorithm}{\protect\numberline{\thealgorithm}##2}%
			\else % #1 is not \relax
			\addcontentsline{loa}{algorithm}{\protect\numberline{\thealgorithm}##1}%
			\fi
			\kern2pt\hrule\kern2pt
		}
	}{% \end{breakablealgorithm}
		\kern2pt\hrule\relax% \@fs@post for \@fs@ruled
	\end{center}
}

\begin{document}
	\title{S2CTrans: Building a Bridge from SPARQL to Cypher\thanks{Zihao Zhao and Xiaodong Ge contribute equally to this paper.}}

	\author{Zihao Zhao\inst{1,2} \and
		Xiaodong Ge\inst{1,2}\and
		Zhihong Shen\inst{1(}\Envelope\inst{)}}
	\authorrunning{Zihao Zhao, Xiaodong Ge et al.}
	\institute{
    Computer Network Information Center, Chinese Academy of Sciences, China \and
    University of Chinese Academy of Sciences, China 
		\email{\{zhaozihao,gexiaodong,bluejoe\}@cnic.cn}}
	
	\maketitle              % typeset the header of the contribution
\begin{abstract}
In graph data applications, data is primarily maintained using two models: RDF (Resource Description Framework) and property graph. The property graph model is widely adopted by industry, leading to property graph databases generally outperforming RDF databases in graph traversal query performance. However, users often prefer SPARQL as their query language, as it is the W3C's recommended standard. Consequently, exploring SPARQL-to-Property-Graph-Query-Language translation is crucial for enhancing graph query language interoperability and enabling effective querying of property graphs using SPARQL. Despite the substantial differences in semantic representation and processing logic between SPARQL and property graph query languages like Cypher, this paper demonstrates the feasibility of translating SPARQL to Cypher for graph traversal queries using graph relational algebra. We present the S2CTrans framework, which achieves SPARQL-to-Cypher translation while preserving the original semantics. Experimental results with the Berlin SPARQL Benchmark (BSBM) datasets show that S2CTrans successfully converts most SELECT queries in the SPARQL 1.1 specification into type-safe Cypher statements, maintaining result consistency and improving the efficiency of data querying using SPARQL.
		
\end{abstract}
	
\section{Introduction}

Knowledge graph models represent the real world through entities, concepts, properties, and their relationships, offering practical and valuable insights for subject research and boasting broad application prospects~\cite{M-Cypher}. Currently, knowledge graph storage primarily relies on two models: Resource Description Framework (RDF)~\cite{RDF1.1} and property graph~\cite{GraphDBModel}. RDF databases, such as Jena~\cite{2006Jena}, maintain the former, while property graph databases, like Neo4j~\cite{Neo4j}, manage the latter.

Owing to its simplicity, intuitiveness, and superior performance, the property graph model has gained widespread adoption in the graph database industry. In general, property graph databases outperform RDF databases in graph traversal and pattern matching tasks. However, users tend to favor SPARQL for data querying, as it is a long-standing W3C recommended standard language. The 2019 W3C Workshop on Web Standardization for Graph Data~\cite{w3c-workshop} called for bridging the gap between RDF and property graph query languages, allowing systems to manage data using the property graph data model while enabling users to query data with SPARQL.

Differences in semantic representation and processing logic exist between SPARQL and the property graph query language, represented by Cypher~\cite{Cypher}, making the standardization process challenging. To promote the standardization of knowledge graph query language and improve interoperability between the Semantic Web and graph database communities, it is necessary to translate SPARQL to Cypher.

The main challenges of this translation are as follows: 
\begin{itemize}
    \item Proving the semantic equivalence of SPARQL and Cypher in graph traversal query.
    \item Resolving the conflict between RDF model and property graph model storage through schema mapping and data mapping.
    \item Designing the pattern matching mapping and solution modifier mapping method to translate SPARQL into Cypher.
\end{itemize}
	
In this study, we establish a graph relational algebra-based semantics for SPARQL and introduce S2CTrans, a provably semantics-preserving SPARQL-to-Cypher translation method. We then evaluate S2CTrans using comprehensive query features on public datasets. Our contributions can be summarized as follows:
	
\begin{itemize}
    \item We establish the semantics of SPARQL based on graph relational algebra, and demonstrate the semantic equivalence between SPARQL and Cypher in representing graph queries.
    \item We introduce the S2CTrans framework, which offers a mapping method for pattern matching and solution modifiers, enabling the translation from SPARQL to Cypher.
    \item We perform a comprehensive query test on large-scale datasets to evaluate the performance improvement of Cypher in graph databases after translating SPARQL using the S2CTrans framework.
\end{itemize}
	
\begin{figure}[h]
    \centering
    \includegraphics[width=\linewidth]{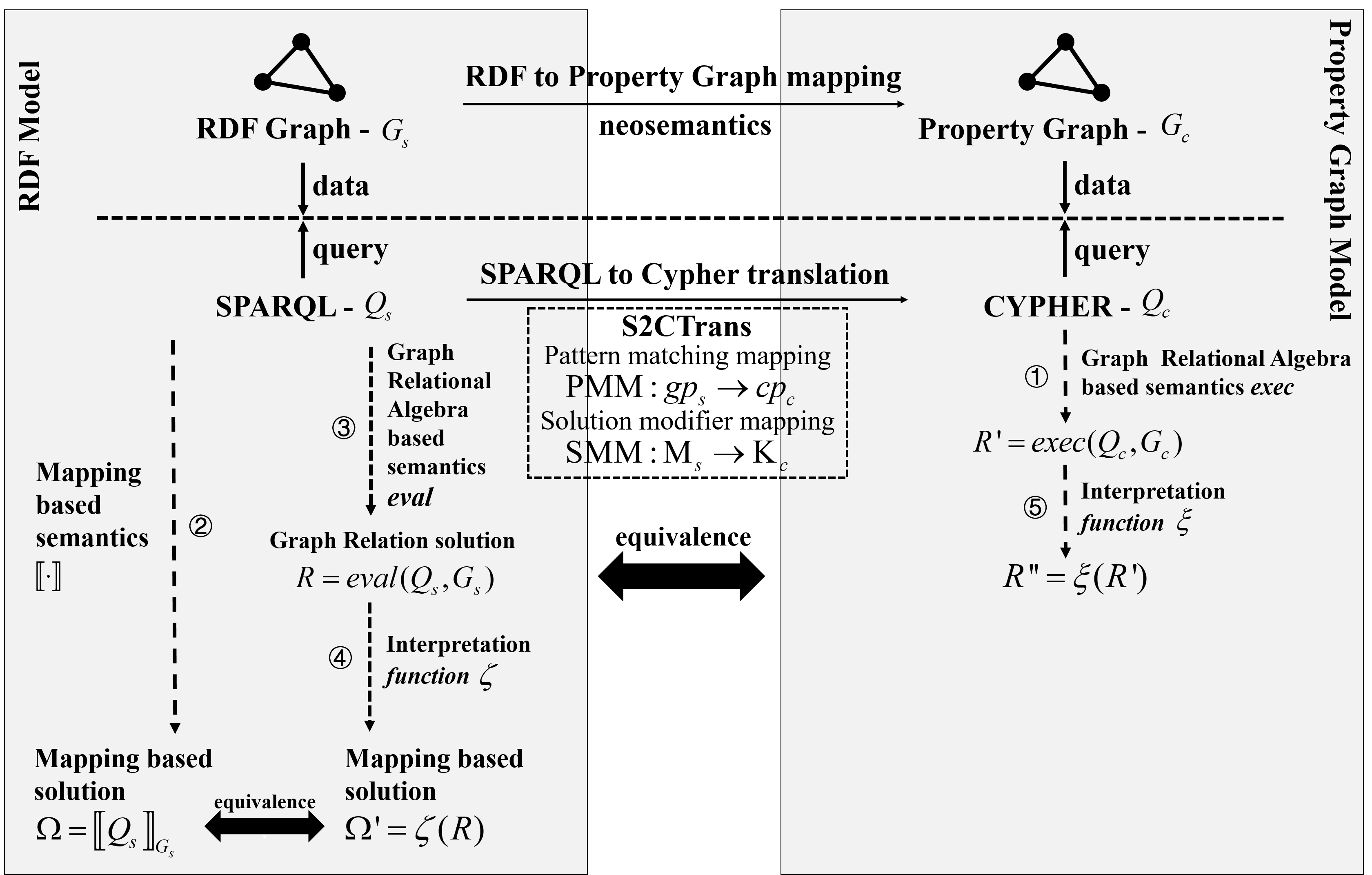}
    \caption{Overview of SPARQL-to-Cypher translation.}\label{fig1}
    \label{fig:overview}
    \vspace{-1em}
\end{figure}

The diagram of our work is illustrated in Figure~\ref{fig:overview}.
At the data level, we implement a syntactic and semantic transformation of RDF graph to property graph using the neosemantincs plug-in \cite{neosemantics} developed by Neo4J Labs. This involves storing RDF triples into property graphs as nodes, relationships, and properties. At the query level, the figure illustrates the first two contributions discussed above. The dashed arrow \ding{172} represents the graph relational algebra of Cypher defined in \cite{graphRelation} \cite{Formalising}, while the dashed arrow \ding{173} represents the mapping-based semantics of SPARQL defined in \cite{Semantics}.Our contributions are represented by the dashed arrows \ding{174}, \ding{175}, and \ding{176}, which define a graph relational algebra based semantics of SPARQL. Additionally, the solid arrows represent our contributions to the definition of the SPARQL-to-Cypher translation, which includes the pattern matching mapping (\textbf{PMM}) and the solution modifier mapping (\textbf{SMM}).
	
The rest of the paper is organized as follows. In Section 2, we present preliminaries for our work. In Section 3, we define a graph relational algebra based semantics of SPARQL and prove the feasibility of SPARQL to Cypher translation. In Section 4, we introduce the system architecture, mapping methods, and limitations of S2CTrans. In Section 5, we present the evaluation strategy, tests, and analyze the experimental results on a large-scale data set. In Section 6, we review related work on the interactivity of the knowledge graph query language. Finally, in Section 7, we conclude the paper and discuss possible future work.
	
\section{Preliminaries}
\subsection{SPARQL Graph Pattern}
Let $I$,$B$,$L$, and $V$ denote pairwise disjoint infinite sets of Internationalized Resource Identifiers (IRIs), blank nodes, literals, and variables, respectively. In the following, we formalize the notions of RDF triple, RDF graph, triple pattern and SPARQL graph pattern.

\begin{definition}
    RDF Triple and RDF Graph. An RDF triple $t$ is a tuple $(s,p,o) \in (I \cup B) \times I \times (I \cup B \cup L)$, where $s$, $p$, and $o$ are a subject, predicate, and object, respectively. An RDF graph $G_s$ is a set of RDF triples.
\end{definition}

\begin{definition}
    Triple Pattern. A triple pattern $tp$ is a triple $(sp, pp, op) \in (I \cup V \cup L) \times (I \cup V) \times (I \cup V \cup L) $, where $sp$, $pp$, and $op$ are a subject pattern, predicate pattern, and object pattern, respectively. 
    % When parsing a triple pattern $tp$, it is necessary to analyze the type of its internal elements $\varphi(x), x \in \{sp,pp,op\}$ and the type of triple pattern $\phi(tp)$. 
    The formal definitions of them are shown in formula (1) and (2) respectively\footnote{In this paper, we only consider the predicate as a variable to represent an unknown edge.} \footnote{We filter the RDF dataset to get the relation type IRI $T$ and property key IRI $P$.}. Table 1 shows the main notations used in this paper.
\end{definition}
% \vspace{-2em}
\begin{equation}
    \varphi(x)=\left\{
    \begin{array}{ll}
        \text{Variable,} & \text{A variable starting with ?.  }  \text{ eg: ?x}\\
        \text{IRI,} & \text{Identify a resource.  } \text{ eg: foaf:know} \\
        \text{Literal,} & \text{Property Value.  }  \text{ eg: 100} \\
        %		\text{Triple,} & \text{Identify a path.  }  \text{ eg: <<?sub bsbm:product ?obj>>}\\
    \end{array}
    \right.
\end{equation}
	
\begin{equation}
    \phi(tp)=\left\{
    \begin{array}{ll}	
        Type, & p=\text{rdf:type / a }\\
        VarEdge, & \varphi(pp)=\text{Variable}  \\
        IRIEdge, &  \varphi(pp)=\text{IRI} \land pp \in T  \\
        Property, & \varphi(pp)=\text{IRI} \land pp \in P
    \end{array}
    \right.
\end{equation}
\vspace{-2em}
\begin{table}[h]
    \caption{List of Notations}
    \resizebox{\textwidth}{21mm}{
        \begin{tabular}{lc|lc}
            \Xcline{1-4}{1pt}
            \textbf{SPARQL Concept} & \textbf{Notation} & \textbf{Cypher Concept} & \textbf{Notation / Set notation}\\
            \Xcline{1-4}{1pt}
            % $G_{RDF}=(V,E)$ & RDF graph\\
            SPARQL query & $Q_s$ & Cypher query & $Q_c$\\
            Triple pattern & $tp=(sp,pp,op)$ & Node patterns & $\chi$\\
            Graph pattern & $gp$ & Relationship patterns & $\rho$   \\
            The type of triple element & $\varphi(x), x \in \{sp,pp,op\}$ & Path patterns & $\epsilon$ / $\mathcal{E}$    \\
            The type of triple pattern & $\phi(tp)$ &  Nodes & $n$ / $\mathcal{N}$\\
            Solution modifier & $\mathbf{M}_s $ & Relationships & $r$ / $\mathcal{R}$  \\
            \Xcline{1-2}{1pt}
            \textbf{S2CTrans Concept} & \textbf{Notation} & Property keys & $p$ / $P$\\ 
            \Xcline{1-2}{1pt}
            Pattern matching mapping  & \textbf{PMM} & Node labels & $l$ / $L$  \\
            Solution modifier mapping & \textbf{SMM} & Relationship types & $t$ / $T$  \\
            Cypher keyword & $\mathbf{K}_c $ & Names & $a$ / $\mathcal{A}$ \\
            - & - & Values & $v$ / $\mathcal{V}$ \\
            \Xcline{1-4}{1pt}
        \end{tabular}
            
    }
\end{table}
\vspace{-2em}
	
	\begin{definition}
		SPARQL Graph Pattern. A SPARQL graph pattern $gp$ is defined by the following abstract grammar:
	\end{definition}
	
	\begin{center}
		$gp \rightarrow tp \ | \ gp \textbf{ AND } gp \ | \ gp \textbf{ OPT } gp \ | \ gp \textbf{ UNION } gp \ | \ gp \textbf{ FILTER } expr_s$
	\end{center}
	where AND, OPT, and UNION are binary operators that correspond to SPARQL conjunction, OPTIONAL, and UNION constructs, respectively. FILTER restricts the solution of graph pattern matching according to the given expression $expr_s$. 
	
%	Due to space constraints, we put the brief definitions of Cypher graph pattern and graph relational algebra in the appendix. For more details, please refer to \cite{Cypher}  \cite{Formalising}.
	
\begin{figure}[h]
    \centering
    \includegraphics[width=\linewidth]{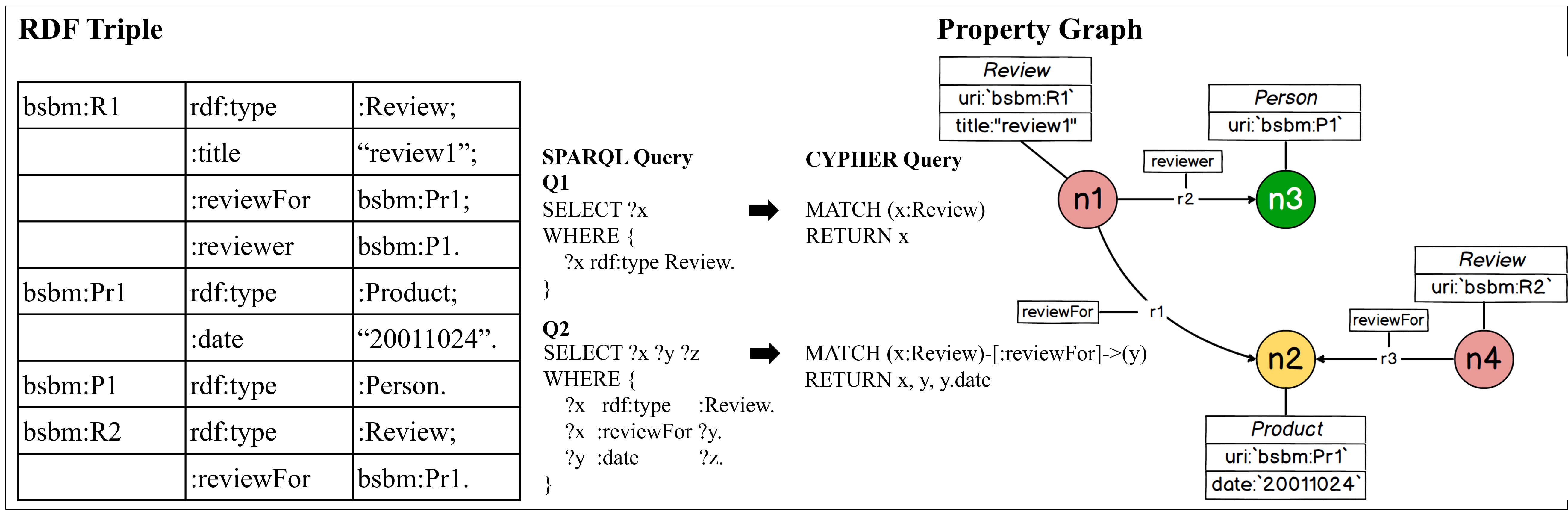}
    \caption{Examples of RDF triples and corresponding property graph, SPARQL queries and corresponding Cypher queries in BSBM dataset.}
    \label{fig2}
\end{figure}
	
\subsection{Cypher Graph Pattern}
\begin{definition}
    Property Graph. A property graph is defined as $G_c = (\mathcal{N},\mathcal{R},st,\mathcal{L}, \mathcal{T},  \mathcal{P})$ where: 
\end{definition}

\begin{itemize}
    \item $\mathcal{N}$ is a set of nodes.
    \item $\mathcal{R}$ is a set of relationships
    \item $st: \mathcal{R} \rightarrow \mathcal{N} \times \mathcal{N}$ assigns the source and target nodes to relationships.
    \item $\mathcal{L}: \mathcal{N} \rightarrow 2^L$ assigns a set of labels to each node.
    \item $\mathcal{T}: \mathcal{R} \rightarrow T$ assigns a single type to each relationship.
    \item $\mathcal{P}: (\mathcal{N} \cup \mathcal{R}) \times P \rightarrow \mathcal{V}$ is a finite partial function that maps a (node or relationship) and a property key to a value.
\end{itemize}
\begin{definition}
    Node Pattern and Relationship Pattern. A node pattern $\chi$ is a triple $(a,L,\mathcal{P}_v)$ where: $a \in \mathcal{A} \cup \{\text{nil}\}$ is an optional name. Taking node $`(n1:Review \ \{uri: \text{`bsbm:R1'}, title: ``review1"\})$' in Fig. 2 as an example, the node pattern of $n1$ is represented as 
\end{definition}

\begin{center}
    $\chi_1 = (n1, \{Review\}, \{uri \mapsto \text{`bsbm:R1'}, title \mapsto ``review1"\})$.
\end{center}

A relationship pattern $\rho$ is a tuple $(d,a,T,\mathcal{P}_e,I)$ where: $d \in \{\rightarrow, \leftarrow, \leftrightarrow\} $ specifies the direction of the pattern. $a \in \mathcal{A} \cup \{\text{nil}\}$ is an optional name. $I$ defines the range of the relationship pattern. Taking relationship $`-[r1:reviewFor]\rightarrow$' in Fig. 2 as an example, the relationship pattern of $r1$ is represented as 
\begin{center}
    $\rho_1 = (\rightarrow, r1, {reviewFor}, \texttt{nil}, \texttt{nil})$.
\end{center}
\begin{definition}
    Cypher Graph Pattern and Combining Graph Pattern. A Cypher graph pattern $\epsilon$ consists of node pattern and path pattern, which is defined by the following abstract grammar:
\end{definition}

\begin{center}
        % \vspace{-1em}
    $\epsilon \rightarrow \chi \ | \ \chi\rho\epsilon$
        % \vspace{-1em}
\end{center}

A \textbf{MATCH} clause defines a graph pattern. A query can be composed of multiple patterns spanning multiple \textbf{MATCH} clauses. A Cypher combining pattern $cp$ is defined by the following abstract grammar \footnote{In Cypher statements, the \textbf{UNION} keyword is used to combine the results of two queries rather than graph patterns.}:
\begin{center}
    $cp \rightarrow \epsilon \ | \ cp \ \textbf{AND} \ cp \ | \ cp \ \textbf{OPT} \ cp \ | \ cp \ \textbf{FILTER} \ expr_c$ 
\end{center}
\subsection{Graph Relational Algebra}
\begin{definition}
Graph Relation. Given a property graph $G_c$, a relation $R$ is a graph relation if the following holds \cite{graphRelation,Formalising}:
\end{definition}
\begin{center}
    $\forall A \in sch(R):dom(A) \subseteq \mathcal{N} \cup \mathcal{R} \cup expr_c$
\end{center}
	
The $sch(R)$ is the schema of $R$, a list containing the attribute names. $dom(A)$ is the domain of attribute $A$. $\mathcal{N}$ and $\mathcal{R}$ are the nodes and relationships of $G_c$, respectively. $expr_c$ represents the expressions of properties, labels, types, and functions in $G_c$.
	
To access a certain property of a node, we use expression $n.p$ to access the corresponding value of property $p$. Also, expression $\mathcal{L}(n)$ returns the labels of node $n$, $\mathcal{T}(r)$ returns the type of relationship $r$, and $\mathcal{P}(n)$ returns the property set of node $n$.
\begin{definition}
    GetNodes Operator. Consider a property graph $G_c$ with a set of nodes $\mathcal{N}$. The $GetNodes$ operator, denoted by $\bigcirc_{(x)}$, returns a graph relation with a single attribute $x$ which contains the nodes of $G_c$ \cite{graphRelation,Formalising}. Taking $Q1$ query in Fig~\ref{fig2} as an example, the corresponding graph relational algebra would be:
\end{definition}

\begin{center}
    $\bigcirc_{(x:Review)}$
\end{center}
\begin{definition}
    Expand Operator. Let $R$ be a graph relation and $x \in sch(R)$ an attribute. The $ExpandOut$ operator $\uparrow_{(x)}^{(y)}[e]$ adds new columns $y$ and $e$ to $R$ containing nodes of $y$ that can be reached by an outgoing relationship $e$ from nodes of $x$. Taking $Q2$ query in Fig. 2 as an example, the corresponding graph relational algebra is shown as follows:
\end{definition}
	
\begin{center}
    $\uparrow_{(x)}^{(y)}[\_r\text{:reviewFor}]\bigcirc_{(x:Review)}$ \footnote{The queried graph pattern might contain anonymous nodes and relationships. In the algebraic form, we denotes this with names starting with an underscore, such as $\_x$ and $\_r$. }
\end{center}

\section{Semantic representation of SPARQL and Cypher}
In this section, we will introduce the semantic-based representation of SPARQL and Cypher according to the numerical order of the dashed arrows in Fig~\ref{fig:overview}.
	
\subsection{Graph relational representation of Cypher query solution.}
Let $exec$ denote a function that defines the graph relational algebra based semantics of S2CTrans generated Cypher statements. This function takes a combining graph pattern $cp$ or a Cypher query $Q_c$ and property graph $G_c$, and returns a graph relation $R$. The definition of $exec$ is presented in Fig~\ref{fig3}(a)\footnote{The high-resolution version of this diagram is included in the appendix S2CTrans-tech-report~\cite{tech-report}.}\cite{Formalising}.

\begin{itemize}
    \item Rule 1 defines the execution of a graph pattern $\epsilon$ over property graph $G_c$. There are two cases:
    \begin{itemize}
        \item $\epsilon$ is a node pattern. Match it with the node set $\mathcal{N}$ in graph $G_c$. If its labels and properties are \texttt{NULL} or consistent with $n,(n \in \mathcal{N})$, then it is said that $n$ satisfies node pattern $\epsilon_{\chi}$, denoted by $(n, G_c) \models \epsilon_{\chi}$.
        
        \item $\epsilon$ is a path pattern. In addition to satisfying the node pattern $\epsilon_{\chi}$, the corresponding relationship pattern $\epsilon_{\rho}$ between nodes needs to be matched with the relationship set $\mathcal{R}$ in $G_c$. If there is a path $p$ in $G_c$, and each sub-relation in $p$ satisfies the corresponding $\epsilon_{\rho}$, then it is said that $n$ and $p$ satisfy the path pattern $\epsilon$, denoted by $(n \cdot p, G_c) \models \epsilon$. 
    \end{itemize}
    Finally, the query results are projected. 
    
    \item Rule 2 defines the execution of the \textbf{AND} of two combining graph patterns $cp_1$ and $cp_2$ as the inner join of graph relations $R_1 = exec(cp_1, G_c)$ and $R_2 = exec(cp_2, G_c)$. 
    
    \item Rule 3 defines the execution of the \textbf{OPT} of two combining graph patterns $cp_1$ and $cp_2$ as the left outer join of graph relations $R_1 = exec(cp_1, G_c)$ and $R_2 = exec(cp_2, G_c)$.
    
    \item Rule 4 defines the execution of the \textbf{UNION} of two queries $Q_{c1}$ and $Q_{c2}$ as the outer union of graph relations $R_1 = exec(Q_{c1}, G_c)$ and $R_2 = exec(Q_{c2}, G_c)$.
    
    \item Rule 5 defines the execution of the \textbf{FILTER} expression $expr_c$ for combining graph pattern $cp$ as the subset of tuples $R$ of graph relations $R_1 = exec(cp, G_c)$. 
\end{itemize}

\begin{sidewaysfigure}
\centering
    \includegraphics[width=\linewidth]{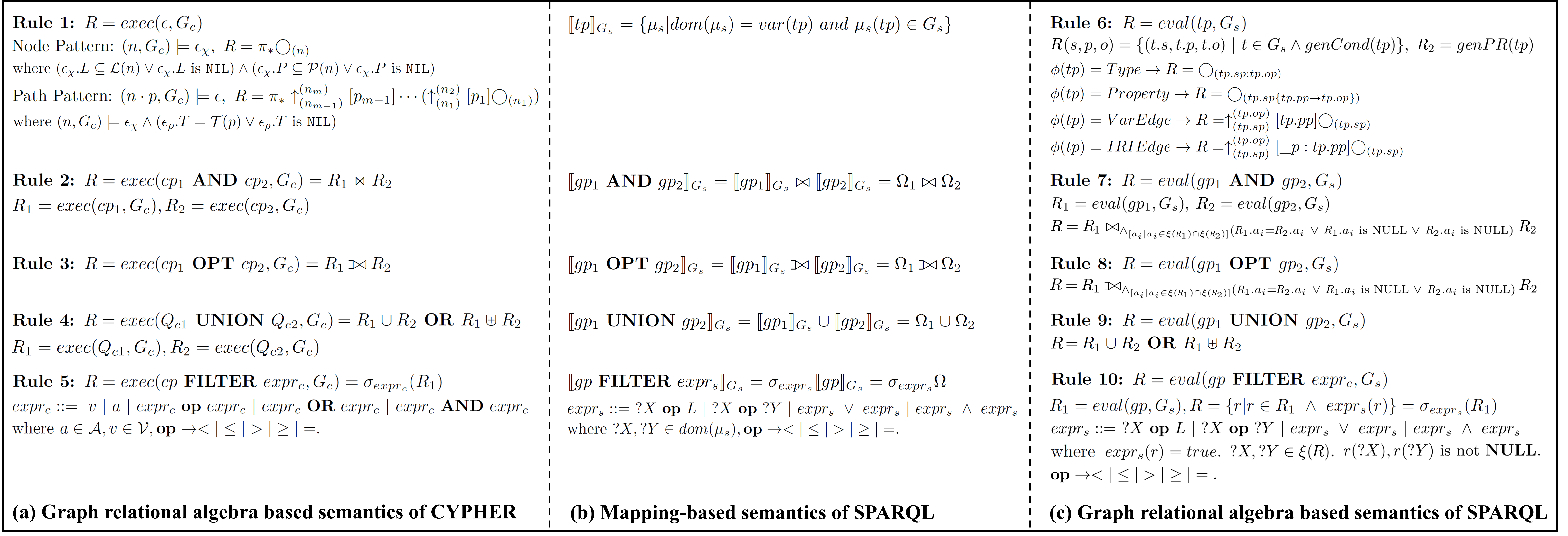} 
    \caption{Semantic-based SPARQL and Cypher solution representation.}
    \label{fig3}
\end{sidewaysfigure}
	
\subsection{Mapping-based representation of a SPARQL query solution.}
	
Let a mapping $\mu : V \rightarrow I \cup B \cup L$ be a partial function that assigns RDF terms to variables of a SPARQL query. The domain of $\mu$, $dom(\mu)$, is the subset of $V$ over which $\mu$ is defined. Then, the mapping-based representation of a SPARQL query solution is a set $\Omega$ of mappings $\mu$. We define $\Sigma$ as an infinite set of all possible mapping-sets, each of which represents a SPARQL query solution.
	
The mapping-based semantics of SPARQL is defined as a functions $[\![ \cdot ]\!]$ which takes a graph pattern expression or a SPARQL query and an RDF graph $G_s$ and returns a set of mappings, denoted as $[\![ \cdot ]\!] : gp \rightarrow \Omega$. The definition of $[\![ \cdot ]\!]$ is presented in Fig. 3(b), where define the evaluation of triple pattern $tp$, $gp_1$ \textbf{AND} $gp_2$, $gp_1$ \textbf{OPT} $gp_2$, $gp_1$ \textbf{UNION} $gp_2$, and $gp$ \textbf{FILTER} $expr_s$, respectively, over an RDF graph $G_s$. Detailed description of $[\![ \cdot ]\!]$ with illustrative examples is available in \cite{Semantics}.
	%
	%SPARQL expressions provide access to operator functions (invoked by keywords and symbols in the SPARQL grammar). Given a mapping $\mu_s$ and expression $expr_s$, $\mu_s$ satisfies $expr_s$, denoted by $\mu_s \models expr_s$. The abstract syntax of the expression $expr_s$ is defined as follows.
	
\subsection{Graph Relational representation of SPARQL query solution.}

Let a tuple $r:IVL \rightarrow IBL \cup \{\texttt{NULL}\}$ be a total function, that assigns RDF terms of an RDF graph to IRIs, literals, and variables of a SPARQL query. The graph relation representation of a SPARQL query solution is a set $R$ of tuples $r$. We define $\mathscr{R}$ as an infinite set of all possible graph relations, each of which represents a SPARQL query solution. 

%The schema of $R$, denoted as $sch(R)$, is the subset of $IVL$ over which each tuple $r \in R$ is defined.We denote a tuple schema as $sch(r)$ and $sch(r) \equiv sch(R)$ for all $r \in R$.

In order to match the triple pattern and modify the query solution, we need to construct the functions $genCond$ and $genPR$. Due to the similarity between Cypher and SQL query matching process, $gencond$ function in  \cite{SematicPreserving} can be reused. Due to the difference in representation between the two solutions, $genPR$ function needs to be rewritten.
	
\begin{breakablealgorithm}  
    \caption{Modify the solution schema based on variables and expression equivalence. }  
    \begin{algorithmic}[1] 
        \Require Triple pattern $tp$
        \Ensure Graph relational algebra expression only projects those attributes of graph relation $R(s, p, o)$ corresponding to the position of variable, node or relation in $tp$, and renames the projected attributes. 
        %		\State \textbf{Input:}  Triples $T$  
        %		\State \textbf{Output:}  Graph Pattern $\epsilon$  
        \Function {genPR}{$tp$}
        \State $project$-$list = s$  
        \State $rename$-$list = s \rightarrow tp.sp$   
        \If {($\phi(tp) = IRIEdge$ OR $\phi(tp) = VarEdge$) AND $tp.pp \neq tp.sp$}
        
        ${project \text{-} list \text{ += } \{p, o\}, rename \text{-} list \text{ += } \{p \rightarrow tp.pp, o \rightarrow tp.op\} }$  
        \EndIf
        
        \If {$\phi(tp) = Type$ AND $\varphi(tp.op) = Variable$}
        
        ${project \text{-} list \text{ += } o, rename \text{-} list \text{ += } o \rightarrow \mathcal{L}(tp.sp)}$
        
        \EndIf
        
        \If {$\phi(tp) = Property$ AND $\varphi(tp.op) = Variable$}
        
        ${project \text{-} list \text{ += } o, rename \text{-} list \text{ += } o \rightarrow (tp.sp).(tp.pp)}$
        
        \EndIf
        \State \Return{$\pi_{project \text{-} list \rightarrow rename \text{-} list}(R)$}  
        \EndFunction  
    \end{algorithmic}  
\end{breakablealgorithm}
	
We define the graph relational algebra based semantics of SPARQL as a function $eval$ which takes a graph pattern expression $gp$ or a SPARQL query $Q_s$ and an RDF graph $G_s$ and returns a graph relation $R$. The definition of $eval$ is presented in Fig~\ref{fig3}(c).
	
\begin{itemize}
    \item Rule 6 defines the evaluation of a triple pattern $tp$ over RDF graph $G_s$ in two steps. First, the graph relation $R$ with the schema $sch(R)=(s,p,o)$ is created and all the triples $t \in G_s$ that match $tp$ based on the condition generated by $genCond(tp)$ are stored into $R$. Then, attributes of $R$ are projected and renamed based on the graph relational algebra expression generated by $genPR(tp)$ and the new graph relation $R_2$ is created. Finally, $R_2$ is assigned as a solution to the triple pattern.
    
    \item Rule 7 defines the evaluation of the \textbf{AND} of two graph patterns $gp_1$ and $gp_2$ as the inner join of graph relations $R_1 = eval(gp_1, G_s)$ and $R_2 = eval(gp_2, G_s)$. The join condition ensures that for every pair of common relational attributes $(R_1.a_i,R_1.a_i)$ where $a_i \in sch(R_1) \cap sch(R_2)$, their values are equal $R_1.a_i = R_2.a_i$ or one or both values are $\texttt{NULLs}$. Finally, the redundant attributes of the join-resulting table are merged into one.
    
    \item Rule 8 defines the evaluation of the \textbf{OPT} of two graph patterns $gp_1$ and $gp_2$ as the left outer join of graph relations $R_1 = eval(gp_1, G_s)$ and $R_2 = eval(gp_2, G_s)$.
    
    \item Rule 9 defines the evaluation of the \textbf{UNION} of two graph patterns $gp_1$ and $gp_2$ as the outer union of graph relations $R_1 = eval(gp_1, G_s)$ and $R_2 = eval(gp_2, G_s)$.
    
    \item Rule 10 defines the evaluation of the \textbf{FILTER} expression $expr_s$ for graph pattern $gp$ as the subset of tuples $R$ of graph relations $R_1 = eval(gp, G_s)$, for which the condition $expr_s(r)$ is true.
\end{itemize}
	
\subsection{Interpretation function}
\subsubsection{Graph relational algebra to Mapping}
Although SPARQL has different solution representations based on graph relational algebra and mapping, both identify a tuple with RDF graph elements. In order to prove their equivalence, we define an interpretation function $\zeta: \mathscr{R} \rightarrow \Sigma$ to relate the graph relation and mapping-based representations. The funtion takes a graph relation $R \in \mathscr{R}$ and returns a mapping-set $\Omega \in \Sigma$, such that each tuple $r \in R$ is assigned a mapping $\mu \in \Omega$ in the following way: if $x \in sch(r), x \in V$ and $r(x)$ is not \texttt{NULL}, then $x \in dom(\mu)$ and $\mu(x) = r(x)$. 
	
Taking query Q2 in Fig~\ref{fig2} as an example, the following mapping shows that the interpretation function $\zeta$ can serve as a tool to establish the equivalence relationship between SPARQL query solutions when different representations are used:
\begin{center}
    $R = \begin{tabular}{cccc}
        \Xcline{1-4}{1pt}
        \textbf{x} & \textbf{:reviewFor} & \textbf{y} & \textbf{y.date}\\
        \Xcline{1-4}{1pt}
        bsbm:R1 & :reviewFor & bsbm:Pr1 & ``20011024''\\
        bsbm:R2 & :reviewFor & bsbm:Pr1 & ``20011024''\\
        \Xcline{1-4}{1pt}
    \end{tabular} \stackrel{\zeta}{\longrightarrow}$
\end{center}

\begin{center}
    $\Omega = \begin{tabular}{lll}
        \Xcline{1-3}{1pt}
        $?x \rightarrow \text{bsbm:R1}$ & $?y \rightarrow \text{bsbm:Pr1}$  & $?z \rightarrow ``20011024$''\\
        \Xcline{1-3}{1pt}
        $?x \rightarrow \text{bsbm:R2}$ & $?y \rightarrow \text{bsbm:Pr1}$  & $?z \rightarrow ``20011024$''\\
        \Xcline{1-3}{1pt}
    \end{tabular}$
\end{center}
	
	\subsubsection{\textit{exec} to \textit{eval}}
	The solution representations of $exec$ and $eval$ are both graph relations. However, due to the storage mechanism of property graph and the transformation strategy of \texttt{neosemantics} \footnote{Neosemantics stores the entity URI in the RDF dataset as a node property in the property graph, so that each node can uniquely identify a node through both node identifier and URI property.}, their schemas are different in representing nodes and relationships. The node identifier and relationship identifier are used in $exec$, while its URI is used in $eval$. Although the two representations are different, both can uniquely identify a node or relationship \footnote{In this paper, we default that there is no duplicate relationship type between any two nodes.}. Therefore, we define an interpretation function $\xi: \mathscr{R} \rightarrow \mathscr{R}$ to modify the schema and tuples of graph relation to explain the equivalence between $exec$ and $eval$. $\xi$ mainly includes the following three steps: 
\begin{itemize}	
    \item Modifying the relationship between the two nodes in the schema to the corresponding relationship URI.
    \item Mapping the node identifier to the corresponding node URI property.
    \item Mapping the relationship identifier to the corresponding relationship type.
\end{itemize}
	
We still take query Q2 in Fig~\ref{fig2} as the example to demonstrate the specific mapping method of interpretation function $\xi$, and prove the equivalence of $eval$ and $exec$.
	
\begin{center}
    $R' = \begin{tabular}{cccc}
        \Xcline{1-4}{1pt}
        \textbf{x} & \textbf{xy} & \textbf{y} & \textbf{y.date}\\
        \Xcline{1-4}{1pt}
        n1 & r1 & n2 & ``20011024''\\
        n4 & r3 & n2 & ``20011024''\\
        \Xcline{1-4}{1pt}
    \end{tabular} \stackrel{\xi}{\longrightarrow}$
\end{center}

\begin{center}
    $R'' = \begin{tabular}{cccc}
        \Xcline{1-4}{1pt}
        \textbf{x} & \textbf{:reviewFor} & \textbf{y} & \textbf{y.date}\\
        \Xcline{1-4}{1pt}
        bsbm:R1 & :reviewFor & bsbm:Pr1 & ``20011024''\\
        bsbm:R2 & :reviewFor & bsbm:Pr1 & ``20011024''\\
        \Xcline{1-4}{1pt}
    \end{tabular} = R$
\end{center}

\section{S2CTrans}
We design and implement S2CTrans, a framework which could equivalently translate SPARQL into Cypher. S2CTrans has been open-sourced\footnote{https://github.com/MaseratiD/S2CTrans}.
\subsection{System Architecture}
\begin{figure}[h]
    \centering
    \includegraphics[width=\linewidth]{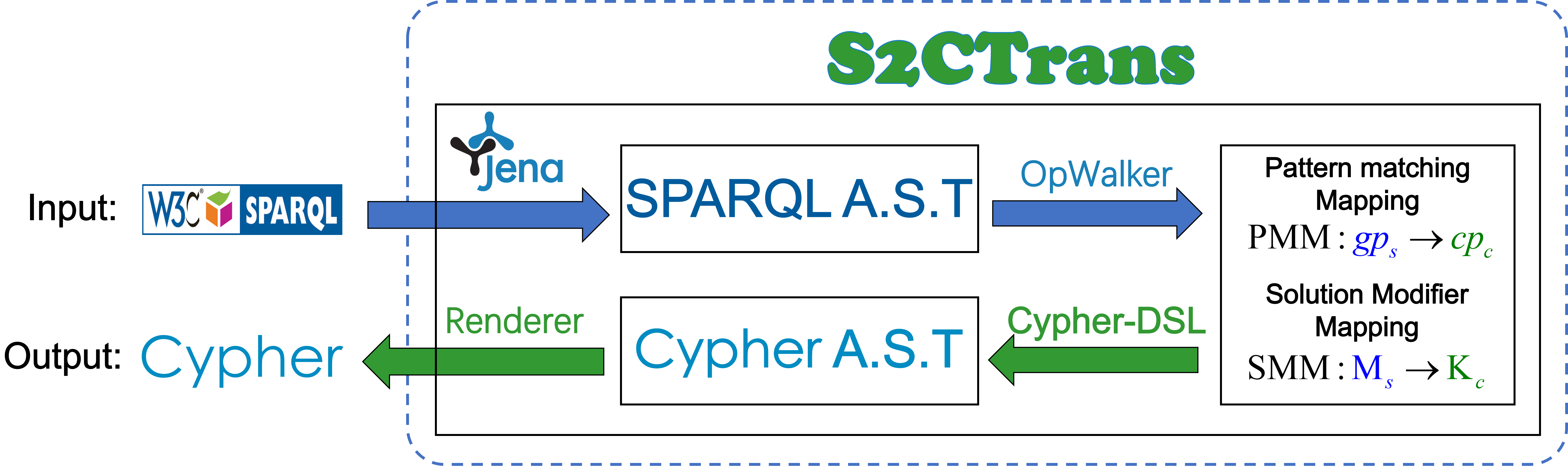}
    \caption{The S2CTrans pipeline architecture.}
    \label{fig4}
    \vspace{-2em}
\end{figure}

S2CTrans takes SPARQL query as input, and generates Cypher statement with the original semantics by using Jena ARQ \cite{2006Jena} parse strategy, graph pattern matching and solution modifiers transformation strategy and Cypher-DSL \cite{Cypher-DSL} construction strategy. We give an overview of the architecture of S2CTrans in Fig~\ref{fig4} and discuss the role of each step in the five-step execution pipeline.
	
\begin{itemize}
    \item \textbf{Step 1:} The input SPARQL query is first parsed by the Jena ARQ module. It can check for syntax errors, verify whether it is a valid SPARQL query and generate an abstract syntax tree (AST) representation.
    \item \textbf{Step 2:} After obtaining the AST parsed by SPARQL, \texttt{OpWalker} is used to access the graph pattern matching part and solution modifier part from bottom up.
    \item \textbf{Step 3:} \textbf{PMM} maps the SPARQL graph pattern $gp_s$ to the Cypher combining graph pattern $cp_c$, and then \textbf{SMM} maps the SPARQL solution modifiers $\mathbf{M}_s$ to Cypher clause keywords $\mathbf{K}_c$.
    \item \textbf{Step 4:} Cypher-DSL generates the final conjunctive traversal and constructs Cypher AST according to the pattern element type and operator priority.
    \item \textbf{Step 5:} Finally, the Cypher AST is rendered as a complete Cypher statement by \texttt{Renderer}. This statement can be directly queried in Neo4j with the \texttt{neosemantics} plug-in to get the result of property graph.
\end{itemize}
	
The mapping function consists of two parts: graph pattern matching mapping \textbf{PMM} and solution modifier mapping \textbf{SMM}.
	
\subsection{Pattern Matching Mapping}
Graph pattern matching is the most basic and important query operation in graph query languages~\cite{Foundations,PatMat}. 
In Table~\ref{tab:example}, we take Fig~\ref{fig2} as an example to show the corresponding mappings from part of the triple patterns $tp$ to Cypher graph pattern $\epsilon$ and Cypher-DSL pattern construction statements. 
Due to page constraints, the graph pattern mapping algorithm is introduced in the appendix. The mapping function \textbf{PMM} in the algorithm translates SPARQL graph pattern into Cypher graph pattern elements. 
	
\begin{table*}[h]
    \label{tab:example}
    \caption{A consolidated list of triple patterns and corresponding Cypher graph patterns.}
    \small
    \resizebox{\textwidth}{17mm}{ 
        \begin{tabular}{ c | c | c | c | c } 
            \Xcline{1-5}{1pt}
            \multicolumn{3}{c|}{\textbf{Triple Pattern - $tp$}} & \multicolumn{2}{c}{\textbf{Cypher Graph Pattern - \textbf{PMM}(tp)}} \\
            \hline
            \textbf{$sp$} & \textbf{$pp$} & \textbf{$op$} & \textbf{Graph Pattern $\epsilon$} & \textbf{Cypher-DSL pattern construction}\\
            \Xcline{1-5}{1pt}
            \textbf{?x} & rdf:type & :Review & $\chi = (x,$ Review$, \varnothing)$ & Cypher.node(``Review'').named(``x'');\\
            \hline
            \textbf{?x} & :title & ``review1'' & $\chi = (x, \varnothing, \{title \mapsto ``review1$''$ \})$ & Cypher.named(``x'').withProperties(``title'',``review1'');\\
            \hline
            \multirow{3}*{\textbf{?x}} & \multirow{3}*{\textbf{reviewFor}} & \multirow{3}*{\textbf{?y} } & $\chi_x = (x, \varnothing, \varnothing)$ & Cypher.anyNode(``x'');\\
            \cline{4-5}
            & & & $\rho = (\rightarrow, \_r,$ reviewFor$, \varnothing, nil)$ & x.relationshipTo(y, ``reviewFor''); \\
            \cline{4-5}
            & & & $\chi_y = (y, \varnothing, \varnothing)$ & Cypher.anyNode(``y'');\\
            \hline
            ?x & rdf:type & \textbf{?y} &  $\chi.getLabel()$ &  Functions.labels(x);\\
            \hline
            \multirow{2}*{?x} & \multirow{2}*{\textbf{?y}} & \multirow{2}*{?z} & $\rho = (\rightarrow, y, \varnothing, \varnothing, nil)$ & x.relationshipTo(z);\\
            \cline{4-5}
            & & & $\rho.getType()$ & Functions.type(y);\\
            \hline
            ?x & :title & \textbf{?y} &  $\chi.getProperty($title$)$ & x.property(``title'');\\
            \hline
            \Xcline{1-5}{1pt}
        \end{tabular}
    }
\end{table*}

	%
	%\begin{table}
	%	\caption{SPARQL and Cypher graph pattern binary operation mapping.}
	%	\begin{tabular}{lll}
	%		\Xcline{1-3}{1pt}
	%		\textbf{Operators} & \textbf{SPARQL} & \textbf{Cypher} \\
	%		\Xcline{1-3}{1pt}
	%		\textbf{AND} & \tabincell{l}{SELECT $varlist$ \\ WHERE  $gp_1$. $gp_2$.  } & \tabincell{l}{MATCH $cp_1$, $cp_2$ \\ RETURN $varlist$} \\
	%		\hline
	%		\textbf{OPT} & \tabincell{l}{SELECT $varlist$ \\ WHERE $gp_1$. \\ OPT $gp_2$.  } & \tabincell{l}{MATCH $cp_1$ \\ OPTIONAL MATCH $cp_2$ \\ RETURN $varlist$} \\
	%		\hline
	%		\textbf{UNION} & \tabincell{l}{SELECT $varlist$ \\ WHERE  $gp_1$. \\ UNION $gp_2$. } & \tabincell{l}{MATCH $cp_1$ RETURN $varlist_1$ \\ UNION \\ MATCH $cp_2$ RETURN $varlist_2$ } \\
	%		\hline
	%		\textbf{FILTER} & \tabincell{l}{SELECT $varlist$ \\ WHERE  $gp$. \\ FILTER $Expr_s$.  } & \tabincell{l}{MATCH $cp$ \\ WHERE $Expr_c$\\ RETURN $varlist$} \\
	%		\Xcline{1-3}{1pt}
	%	\end{tabular}
	%\end{table}

\subsection{Solution Modifiers Mapping}
After the graph pattern is obtained by \textbf{PMM} algorithm, conditions are usually added to modify the solution of graph pattern matching. Based on the semantic equivalence of SPARQL and Cypher in graph relational algebraic expressions, \textbf{SMM} algorithm constructs a mapping table (as shown in Table~\ref{tab:smm}) to implement the mapping of SPARQL solution modifiers $\mathbf{M}_s$ to Cypher clause keywords $\mathbf{K}_c$. This table summarizes graph query modification operations and the corresponding graph relational algebra, as well as the forms of SPARQL and Cypher clause construction. The variables and expressions have been mapped to graph pattern elements in \textbf{PMM} algorithm.
	
\begin{table*}[h]
    \caption{A consolidated list of SPARQL solution modifiers and corresponding Cypher clause keywords.}
    \label{tab:smm}
    \small
    \resizebox{\textwidth}{10mm}{
        \begin{tabular}{l|c|l|l}
            \Xcline{1-4}{1pt}
            \textbf{Operation} & \textbf{Algebra} & \textbf{SPARQL Solution Modifiers - $\mathbf{M}_s$} & \textbf{Cypher Clause Keywords - $\mathbf{K}_c$}\\ 
            \Xcline{1-4}{1pt}
            Selection & $\sigma_{condition}(r)$ & FILTER($Expr_1$ $\&\& (||)$ $Expr_2$) & WHERE $Expr_1$ $ and(or)$ $Expr_2$\\ \hline
            Projection & $\pi_{x_1,x_2,...}(r)$ & SELECT $?x_1$ $?x_2$ ...  & RETURN $x_1$, $x_2$, ...\\ \hline
            De-duplication & $\delta_{x_1,x_2,...}(r)$ & SELECT DISTINCT $?x_1$ $?x_2$ ...  & RETURN DISTINCT $x_1$, $x_2$, ...\\ \hline
            Restriction & $\lambda_{s}^{l}(r)$ & LIMIT $l$ SKIP $s$ & LIMIT $l$ SKIP $s$\\ \hline
            %		Grouping & $\gamma_{x_1,x_2,...}^{c_1,c_2,...}(r)$ & GROUP BY( $?x$ ) & -\\ \hline
            Sorting & $\varsigma_{\uparrow_{x_1}, \downarrow_{x_2}, ...}(r)$ & ORDER BY ASC($?x_1$) DESC($?x_2$) & ORDER BY $x_1$ ASC, $x_2$ DESC\\ 
            \Xcline{1-4}{1pt}
        \end{tabular}
    }
\end{table*}

Through \textbf{PMM} algorithm and \textbf{SMM} algorithm, we get the Cypher graph pattern and clause keywords. Cypher-DSL constructs Cypher AST according to graph pattern elements and operator precedence. Finally, we use \texttt{Renderer} to construct a complete Cypher statement.
	
\section{Experiments}
\subsection{Evaluation criteria}
We execute SPARQL queries on several top-of-the-line RDF databases, and execute translated Cypher queries on graph database Neo4j. We evaluate S2CTrans by the translation speed, query execution time and result consistency.

\subsection{Experimental setup}
\subsubsection{\textbf{Dataset}}
This experiment uses the Berlin SPARQL Benchmark(BSBM) dataset recommended by W3C, which consists of synthetic data describing e-commerce use cases, involving categories such as products, producers, etc. We generated 10M triples respectively by BSBM-Tools, and the corresponding property graph version is mapped using the neosemantics plug-in. The details of dataset are introduced in the appendix.
	
\subsubsection{\textbf{Query statements}}
We created a total of 40 SPARQL queries, covering 30 different query features. These queries were selected after systematically studying the semantics of SPARQL queries~\cite{Semantics}. The queries are detailed in the appendix.
\subsubsection{\textbf{System Setup}}
We execute the query statements on the following databases to evaluate the performance improvement of S2CTrans:
\textbf{Property Graph Database:} Neo4j v4.2.3
\textbf{RDF Databases:} Virtuoso v7.2.5, Stardog v7.6.3, RDF4J v3.6.3, Jena TDB v4.0.0
All experiments were performed on the following machine configurations: CPU: Intel Core Processor (Haswell) 2.1GHz; RAM: 16 GB DDR4; HDD: 512 GB SSD; OS: CentOS 7. 
In order to ensure the reproducibility of the experimental results, we provide the experimental script, dataset and query statement\footnote{https://github.com/MaseratiD/S2CTrans}. 
	
\subsection{Result Evaluation}
\begin{figure*}[h]
    \centering
    \includegraphics[width=\linewidth]{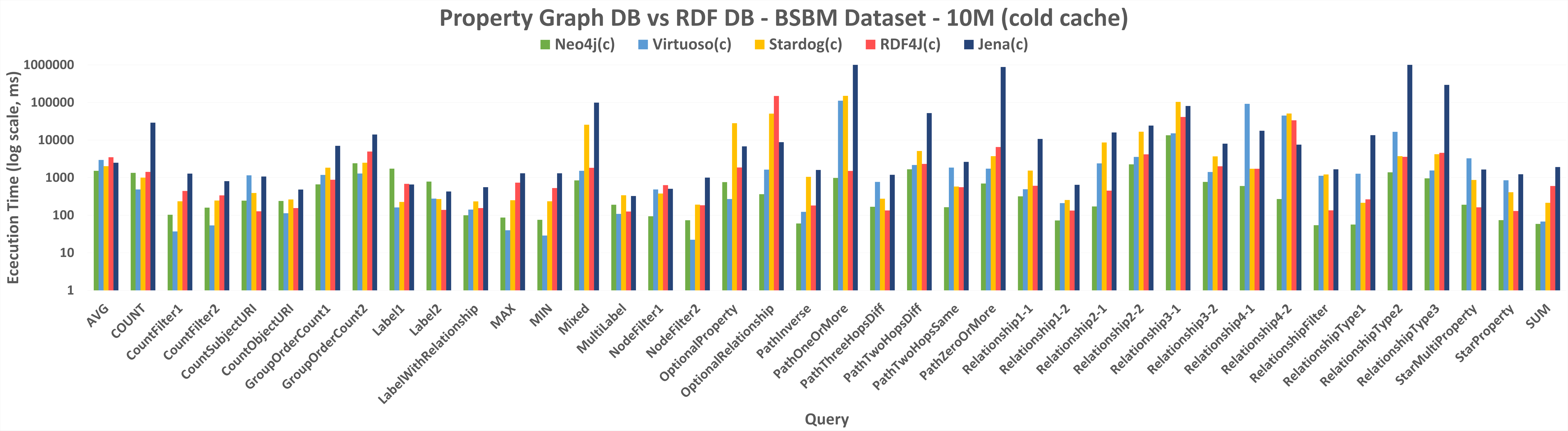}
    \caption{Property graph database V.S. RDF database - BSBM Dataset\_10M}
    \label{fig5}
    %	\Description{Property DBMS vs RDF DBMS - BSBM Dataset - 10M(cold cache)}
\end{figure*}

According to the evaluation criteria described above, we perform SPARQL query on RDF databases and the translated Cypher query on property graph database Neo4j on the dataset. Finally, we compare and analyze the query results. Among them, each query runs an average of 10 times to get the average value. Due to the limited space of the paper, the statements translations and query results are shown in the appendix of S2C-tech-report~\cite{tech-report}.
\begin{itemize}
    \item \textbf{Consistency:} In all experiments, the query results of SPARQL and the Cypher obtained through S2CTrans translation were completely identical. This demonstrates that S2CTrans can equivalently translate SPARQL query statements into Cypher query statements.
    \item \textbf{Performance Analysis: }
    \begin{itemize}
        \item \textbf{Translation Time:} The average translation time of S2CTrans of 40 queries on BSBM-10M is \textbf{23.7ms}. Compared with the query time, it accounts for a small proportion. 
        \item \textbf{Query Performance:} We meticulously conducted tests on datasets of various scales under both cold-start and warm-start scenarios, and all tests yielded similar results. Figure~\ref{fig5} presents the query execution time during the system's cold-start phase. Other test results are detailed in the appendix.
        Among most query statements, Neo4j performs better than the RDF databases. Moreover, in the queries with multi-hop paths and long relationships, the performance of Neo4j is 1 to 2 orders of magnitude higher than RDF database. The main reason is that RDF database spends a lot of time in executing join operation and forming execution plan, while Neo4j uses index-free adjacency, which greatly improves the query efficiency. 
        % However, in the query node type statement, because the index cannot be used, Neo4j runs for a long time.
        %		\item \textbf{Query Volatility:} The average query fluctuation degree of the graph database Neo4j is about 100 milliseconds, while the RDF database is roughly 1 to 2 orders of magnitude higher. This shows that graph database query is less volatile and more stable than the RDF database.
        %		\item \textbf{Query Scalability:} When the size of the dataset is expanded 10 times, the query execution time of the graph database Neo4j increases by 3 times on average, while RDF database increases by 10 times or more on average. This shows that graph databases have better query scalability than RDF databases.
    \end{itemize}
\end{itemize}
The experiment results prove that the proposed S2CTrans is successful in equivalent translating and executing SPARQL queries. S2CTrans enables the users to query property graph by SPARQL.
%\begin{figure*}[h]
%	\centering
%	\includegraphics[width=\linewidth]{10Mwarm-2.png}
%	\caption{Property graph database V.S. RDF database - BSBM Dataset\_10M(warm cache)}\label{fig6}
%	%	\Description{Property DBMS vs RDF DBMS - BSBM Dataset - 10M(warm cache)}
%\end{figure*}

% \begin{figure*}[h]
%   \centering
%   \includegraphics[width=\linewidth]{scaleIncrease.png}
%   \caption{Query stability comparison (Data scale expanded from 1M to 10M)}
%   \Description{Query stability comparison (Data scale expanded from 1M to 10M)}
% \end{figure*}

\section{Conclusion}
In this paper, we first establish the feasibility of translating SPARQL to Cypher based on graph relational algebra and mapping semantics. We then introduce S2CTrans, a novel approach that supports SPARQL-to-Cypher translation. This method can convert most SPARQL statements into type-safe Cypher statements. Moreover, we employ property graph databases and RDF databases to conduct experimental evaluations on large-scale datasets, validating the effectiveness and applicability of our approach. The evaluation highlights the substantial performance gains achieved by translating SPARQL queries to Cypher queries, particularly for multiple relationship and star-shaped queries. Although S2CTrans currently has several limitations, it represents an important step toward promoting the standardization of graph query languages and enhancing the interoperability of data and queries between the Semantic Web and graph database communities. In the future, we plan to further refine S2CTrans to support more SPARQL translations and investigate the translation from Cypher to SPARQL.

\section{Appendix}
	
\subsection{Graph Pattern Mapping Algorithm}
    \begin{breakablealgorithm}  
        \caption{Pattern matching mapping}  
        \begin{algorithmic}[1] 
            \Require SPARQL graph pattern $gp$
            \Ensure Cypher Graph Pattern Set $\mathcal{E}$ 
            %		\State \textbf{Input:}  Triples $T$  
            %		\State \textbf{Output:}  Graph Pattern $\epsilon$  
            \Function {PMM}{$gp$}
            \State $\mathcal{E} \gets \varnothing$  
            \For {$\text{triple pattern } tp \in gp, tp=(sp,pp,op)$} 
            \State $\epsilon \gets {\mathcal{E}}.getPattern(sp)$  
            \If{$\epsilon = \varnothing$}
            \IIf {$\phi(tp) = Type$}  $\epsilon = \chi = (sp, op, nil)$  \EndIIf
            \IIf {$\phi(tp) = Property$} $\epsilon = \chi = (sp, \varnothing, {pp \mapsto op})$ \EndIIf
            
            \State $\chi_s \gets (sp, \varnothing, nil)$, $\chi_o \gets (op, \varnothing, nil)$
            \IIf {$\phi(tp) = IRIEdge$} $\rho = (\rightarrow, \_p, pp, \varnothing, nil)$, $\epsilon = \chi_s\rho\chi_o$ \EndIIf
            \IIf {$\phi(tp) = VarEdge$} $\rho = (\rightarrow, pp, \varnothing, \varnothing, nil)$, $\epsilon = \chi_s\rho\chi_o$  \EndIIf
            \ElsIf {$\epsilon$ is NodePattern} 
            \IIf {$\phi(tp) = Type$}  $\epsilon = {\epsilon}.addLabel(op)$  \EndIIf
            \IIf {$\phi(tp) = Property$}  $\epsilon = {\epsilon}.addProperty({pp \mapsto op})$ \EndIIf
            
            \State $\chi_o \gets (op, \varnothing, nil)$
            \IIf {$\phi(tp) = IRIEdge$} $\rho = (\rightarrow, \_p, pp, \varnothing, nil)$,  $\epsilon = \epsilon\rho\chi_o$ \EndIIf
            \IIf {$\phi(tp) = VarEdge$} $\rho = (\rightarrow, pp, \varnothing, \varnothing, nil)$,  $\epsilon = \epsilon\rho\chi_o$ \EndIIf
            \Else
            \State $\chi_1 \gets {\epsilon}.getStartNode()$, $\rho \gets {\epsilon}.getRelation() $, $\chi_2 \gets {\epsilon}.getEndNode()$
            \State $\chi \gets sp = {\chi_1}.getName() \text{ ? } \chi_1 : \chi_2 $
            \IIf {$\phi(tp) = Type$}  $\epsilon = {\chi}.addLabel(op)$  \EndIIf
            \IIf {$\phi(tp) = Property$}  $\epsilon = {\chi}.addProperty({pp \mapsto op})$ \EndIIf
            \State $\chi_3 \gets (op, \varnothing, nil)$
            \If{$\chi = \chi_1$}
            \IIf {$\phi(tp) = IRIEdge$} $\rho' = (\leftarrow, \_p, pp, \varnothing, nil)$,  $\epsilon = \chi_3\rho'\chi\rho\chi_2$ \EndIIf
            \IIf {$\phi(tp) = VarEdge$} $\rho' = (\leftarrow, pp, \varnothing, \varnothing, nil)$,  $\epsilon = \chi_3\rho'\chi\rho\chi_2$ \EndIIf
            \Else
            \IIf {$\phi(tp) = IRIEdge$} $\rho' = (\rightarrow, \_p, pp, \varnothing, nil)$,  $\epsilon = \chi_1\rho\chi\rho'\chi_3$ \EndIIf
            \IIf {$\phi(tp) = VarEdge$} $\rho' = (\rightarrow, pp, \varnothing, \varnothing, nil)$,  $\epsilon = \chi_1\rho\chi\rho'\chi_3$ \EndIIf
            \EndIf
            \EndIf
            %		\State $\epsilon = \chi_1\rho\chi_2$
            \State ${\mathcal{E}}.addPattern(\epsilon)$
            \EndFor
            \State \Return{$\mathcal{E}$}  
            \EndFunction  
        \end{algorithmic}  
    \end{breakablealgorithm}

\subsection{SPARQL Property Path Translation}
For the translation of property paths in SPARQL1.1, we first extract the path identifier of the predicate in the triple pattern, and construct the corresponding Cypher relationship pattern with the strategy in Table 4 and add it to the Cypher combining graph pattern. Finally, pattern matching is performed on the property graph.
\begin{table}[h]
    \caption{SPARQL Property Path Translate to Cypher Path Pattern}
    \resizebox{\textwidth}{17mm}{
        \begin{tabular}{lccl}
            \toprule[1pt]
            \textbf{Path Type} & \textbf{Triples} & \textbf{Graph Pattern} & \textbf{Explanation} \\
            \midrule[1pt]
            PredicatePath & ?s iri:rel ?o .  & $\rho = (\rightarrow, rel',$ iri:rel$, \varnothing, nil)$ & A path of length one.\\
            \hline
            InversePath & ?s \^{}iri:rel ?o . & $\rho = (\leftarrow, rel',$ iri:rel$, \varnothing, nil)$ & Inverse path (object to subject).\\
            \hline
            ZeroOrMorePath & ?s iri:rel* ?o . & $\rho = (\rightarrow, rel',$ iri:rel$, \varnothing, (0, nil))$ & A path of length zero or more.\\
            \hline
            OneOrMorePath & ?s iri:rel+ ?o . & $\rho = (\rightarrow, rel',$ iri:rel$, \varnothing, (1, nil))$ & A path of length one or more. \\
            \hline
            ZeroOrOnePath & ?s iri:rel? ?o . & $\rho = (\rightarrow, rel',$ iri:rel$, \varnothing, (0, 1))$ & A path of length zero or one. \\
            \hline
            SequencePath & ?s iri:rel1/iri:rel2  ?o . & \tabincell{c}{$\rho_1 = (\rightarrow, rel1',$ iri:rel1$, \varnothing, nil)$ \\ $\rho_2 = (\rightarrow, rel2',$ iri:rel2$, \varnothing, nil)$ \\ $\epsilon = \chi_s \rho_1 \chi \rho_2 \chi_o$ } & \tabincell{l}{A sequence path of rel1 followed \\ by rel2.}\\
            \bottomrule[1pt]
        \end{tabular}
    }
\end{table}

\begin{table}
    \centering
    \caption{SPARQL and Cypher graph pattern binary operation mapping.}
    \begin{tabular}{lll}
        \Xcline{1-3}{1pt}
        \textbf{Operators} & \textbf{SPARQL} & \textbf{Cypher} \\
        \Xcline{1-3}{1pt}
        \textbf{AND} & \tabincell{l}{SELECT $varlist$ \\ WHERE  $gp_1$. $gp_2$.  } & \tabincell{l}{MATCH $cp_1$, $cp_2$ \\ RETURN $varlist$} \\
        \hline
        \textbf{OPT} & \tabincell{l}{SELECT $varlist$ \\ WHERE $gp_1$. \\ OPT $gp_2$.  } & \tabincell{l}{MATCH $cp_1$ \\ OPTIONAL MATCH $cp_2$ \\ RETURN $varlist$} \\
        \hline
        \textbf{UNION} & \tabincell{l}{SELECT $varlist$ \\ WHERE  $gp_1$. \\ UNION $gp_2$. } & \tabincell{l}{MATCH $cp_1$ RETURN $varlist_1$ \\ UNION \\ MATCH $cp_2$ RETURN $varlist_2$ } \\
        \hline
        \textbf{FILTER} & \tabincell{l}{SELECT $varlist$ \\ WHERE  $gp$. \\ FILTER $Expr_s$.  } & \tabincell{l}{MATCH $cp$ \\ WHERE $Expr_c$\\ RETURN $varlist$} \\
        \Xcline{1-3}{1pt}
    \end{tabular}
\end{table}

\subsection{Query Feature}
The query feature component of the experimental part is shown in Table 6.
\begin{table*}[h]
    \caption{List of Query Feature Component}
    \resizebox{\textwidth}{72mm}{
        \begin{tabular}{lcccccccc}
            \toprule
            \textbf{Query} & \textbf{Aggregator} & \textbf{Filters} & \textbf{Order} & \textbf{Distinct} & \textbf{Limit} & \textbf{Optional} & \textbf{\#Tps} & \textbf{\#Projs}\\
            \midrule
            COUNT & COUNT &  &  &  &  & & 2 & 1  \\
            CountFilter1 & COUNT & $\surd$(1) &  &  &  &  & 3 & 1  \\
            CountFilter2 & COUNT & $\surd$(2) &  &  &  &  & 4 & 1  \\
            CountSubjectURI & COUNT &  &  &  &  &  & 3 & 1  \\
            CountObjectURI & COUNT &  &  &  &  &  & 3 & 1  \\
            MAX & MAX & $\surd$(1) &  &  &  &  & 3 & 1 \\
            MIN & MIN & $\surd$(1) &  &  &  &  & 3 & 1 \\
            SUM & SUM & $\surd$(1) &  &  &  &  & 3 & 1 \\
            AVG & AVG & $\surd$(1) &  &  &  &  & 3 & 1 \\
            GroupOrder1 & COUNT & $\surd$(1) & $\surd$(1) & $\surd$  & $\surd$ &  & 7 & 2 \\
            GroupOrder2 & COUNT &  & $\surd$(1) & $\surd$  & $\surd$ &  & 4 & 2 \\
            NodeFilter1 &  & $\surd$(1) &  & $\surd$  & $\surd$ &  & 4 & 1\\
            NodeFilter2 &  & $\surd$(2) &  & $\surd$  & $\surd$ &  & 4 & 1\\
            RelationshipFilter &  & $\surd$(1) &  & $\surd$  & $\surd$ &  & 6 & 1\\
            Label1 &  &  &  &   & $\surd$ &  & 2 & 1\\
            Label2 &  &  &  &   & $\surd$ &  & 3 & 2\\
            LabelWithRelationship &  &  &  &   & $\surd$ &  & 4 & 1\\
            MultiLabel &  & $\surd$(1) & $\surd$(2) & $\surd$ & $\surd$ &  & 5 & 1\\
            PathInverse &  & $\surd$(1) & $\surd$(1) & $\surd$ & $\surd$ &  & 6 & 1\\
            PathOneOrMore &  & $\surd$(1) & $\surd$(1) & $\surd$ & $\surd$ &  & 6 & 1\\
            PathZeroOrMore &  & $\surd$(1) & $\surd$(1) & $\surd$ & $\surd$ &  & 5 & 1\\
            PathTwoHopsSame &  & $\surd$(1) & $\surd$(1) & $\surd$ & $\surd$ &  & 5 & 1\\
            PathTwoHopsDiff &  &  & $\surd$(1) & $\surd$ & $\surd$ &  & 5 & 2\\
            PathThreeHopsDiff &  &  &  & $\surd$ & $\surd$ &  & 3 & 1\\
            Relationship1-1 &  & $\surd$(1) & $\surd$(1) & $\surd$ & $\surd$ &  & 5 & 1\\
            Relationship1-2 &  & $\surd$(2) & $\surd$(2) & $\surd$ & $\surd$ &  & 5 & 2\\
            Relationship2-1 &  & $\surd$(1) & $\surd$(2) & $\surd$ & $\surd$ &  & 8 & 2\\
            Relationship2-2 &  &  & $\surd$(2) & $\surd$ & $\surd$ &  & 6 & 2\\
            Relationship3-1 &  & $\surd$(1) & $\surd$(2) & $\surd$ & $\surd$ &  & 9 & 2\\
            Relationship3-2 &  & $\surd$(2) & $\surd$(2) & $\surd$ & $\surd$ &  & 9 & 2\\
            Relationship4-1 &  & $\surd$(1) & $\surd$(2) & $\surd$ & $\surd$ &  & 10 & 2\\
            Relationship4-2 &  & $\surd$(2) & $\surd$(2) & $\surd$ & $\surd$ &  & 10 & 2\\
            RelationshipType1 &  &  &  & $\surd$ & $\surd$ & & 4 & 1 \\
            RelationshipType2 &  &  &  & $\surd$ & $\surd$ & & 3 & 2\\
            RelationshipType3 &  &  &  & $\surd$ & $\surd$ & & 3 & 1\\
            StarMultiProperty &  &  & $\surd$(1) & $\surd$ & $\surd$ &  & 14 & 7\\
            StarProperty &  & $\surd$(1) & $\surd$(1) & $\surd$ & $\surd$ &  & 6 & 1\\
            OptionalProperty &  & $\surd$(1) & $\surd$(1) & $\surd$ & $\surd$ & $\surd$ & 6 & 1 \\
            OptionalRelationship &  & $\surd$(1) & $\surd$(1) & $\surd$ & $\surd$ & $\surd$ & 7 & 1\\
            Mixed & MAX & $\surd$(1) & $\surd$(1) & $\surd$ & $\surd$ & $\surd$ & 6 & 2\\
            \bottomrule
            \textbf{Total} & \textbf{40} & - & - & - & - & - & - & -\\
            \bottomrule
        \end{tabular}
    }
\end{table*}

\subsection{Query Statements and Results}
% Dataset Description
\begin{table}
    \centering
    \caption{Dataset Description}
    \begin{tabular}{lcc}
        \toprule
        & \textbf{BSBM-1M} & \textbf{BSBM-10M} \\
        \midrule  
        \textbf{RDF Triples} & 1,000,313 & 10,031,929\\
        \textbf{PG Nodes} & 148,542 & 1,479,142\\
        \textbf{PG Relationships} & 377,201 & 3,821,818\\
        \textbf{PG Labels} & 160 & 594\\
        \bottomrule
    \end{tabular}
\end{table}

Table 7 shows part of the query translation results and query execution results of the experiment.
\begin{table*}[h]	
    \caption{Query example}
    \scriptsize
    \begin{adjustbox}{angle=270}
        \resizebox{\textheight}{65mm}{
            \begin{tabular}{p{0.1\linewidth}  p{0.25\linewidth}  p{0.34\linewidth}  p{0.4\linewidth}  p{0.35\linewidth}}
                \toprule
                \textbf{QueryNo.} & \textbf{SPARQL} & \textbf{Cypher} & \textbf{SPARQL Result} & \textbf{Cypher Result}\\
                \midrule  
                Count1 & 
                
                SELECT (count(?p) as ?total)
                
                WHERE\{    
                
                \ \  ?R a b: R.    
                
                \ \  ?R b: rF ?p. 
                
                \} & 
                
                MATCH (R:`b:R`)-[rF:`b:rF`]$\rightarrow$(p) 
                
                RETURN count(p) AS total & --------------
                
                $|$ total $|$
                
                ======
                
                $|$ 283000 $|$
                
                --------------
                & [total: 283000] \\
                
                \midrule 
                NodeFil-er2 & 
                
                SELECT DISTINCT ?p2 
                
                WHERE \{ 	
                
                \ \  ?p a b-inst:PT1.	
                
                \ \  ?p b: pPN1 ?p1.    
                
                \ \  ?p b: pPN2 ?p2.    
                
                \ \  FILTER ( ?p1 = 1 ) 
                
                \}
                
                LIMIT 3 &
                
                MATCH (p:`b-inst:PT1`) 
                
                WHERE p.`b:pPN1` = 1 
                
                RETURN DISTINCT p.`b:pPN2` 
                
                LIMIT 3 & 
                ---------------------
                
                $|$ p2 $|$
                
                =========
                
                $|$ 330       $|$
                
                $|$ 246       $|$
                
                $|$ 584       $|$
                
                --------------------- &
                [R.`b:pPN2`: 330]
                
                [R.`b:pPN2`: 584]
                
                [R.`b:pPN2`: 246]
                
                \\
                \midrule
                Relaion-ship 1-2 & 
                SELECT DISTINCT ?pPN1 ?label 
                
                WHERE\{     
                
                \ \ ?p rdf:type b-inst:PT80. 	
                
                \ \ ?p b:pPN1 ?pPN1.	
                
                \ \ ?p b:pr ?pr1.	
                
                \ \ ?pr1 rdf:type b:pr.	
                
                \ \ ?pr1 rdfs:label ?label.	
                
                \ \ FILTER(?pPN1 $<$ 300 \&\& ?pPN1 $>$ 100)
                
                \}
                
                ORDER BY(?label) DESC(?pPN1) 
                
                LIMIT 5 & 
                
                MATCH (p:`b-inst:PT80`)-[pr:`b:pr`]$\rightarrow$(pr1:`b:pr`) 
                
                WHERE (p.`b: pPN1` $<$ 300 AND p.`b: pPN1` $>$ 100) 
                
                RETURN DISTINCT p.`b: pPN1`, pr1.`rdfs: label` 
                
                ORDER BY pr1.`rdfs: label` ASC, p.`b: pPN1` DESC 
                
                LIMIT 5 & 
                
                ---------------------------------------------------

                $|$ pPN1 $|$ label                                $|$
                
                ======================
                
                $|$ 107  $|$ ``ascendence doses whitecaps''      $|$
                
                $|$ 135  $|$ ``ashrams winnings graving''        $|$
                
                $|$ 245  $|$ ``cellulosic''                      $|$
                
                $|$ 163  $|$ ``counterpane menthol harries''     $|$
                
                $|$ 186  $|$ ``futilely''                        $|$
                
                ---------------------------------------------------
                & 
                
                [p.`b:pPN1`:107, pr1.`rdfs:label`: ``ascendence doses whitecaps'']
                
                [p.`b:pPN1`:135, pr1.`rdfs:label`: ``ashrams winnings graving'']
                
                [p.`b:pPN1`:245, pr1.`rdfs:label`: ``cellulosic'']
                
                [p.`b:pPN1`:163, pr1.`rdfs:label`: ``counterpane menthol harries'']
                
                [p.`b:pPN1`:186, pr1.`rdfs:label`: ``futilely''] 
                \\
                \midrule
                RelType3 &  
                SELECT DISTINCT ?v 
                
                WHERE \{ 	
                
                \ \ ?r a b:O. 	
                
                \ \ ?r ?v ?v1.	
                
                \ \ ?v1 a b:V.
                
                \} 
                
                LIMIT 10 & 
                
                MATCH (r:`b:O`)-[v]$\rightarrow$(v1:`bs-bm:V`) 
                
                RETURN DISTINCT type(v) 
                
                LIMIT 10 &
                
                -----------------------------------------
                
                $|$ v                        $|$
                
                =================
                
                $|<$http://XXX.publisher$>                      |$
                
                $|<$http://XXX.v$> |$
                
                ------------------------------------------ & 
                
                [type(v): ``http://XXX.publisher'']
                
                [type(v): ``http://XXX.v''] \\
                \bottomrule
            \end{tabular}.
        }
    \end{adjustbox}
\end{table*}

\begin{sidewaysfigure}
\centering
    \includegraphics[width=\linewidth]{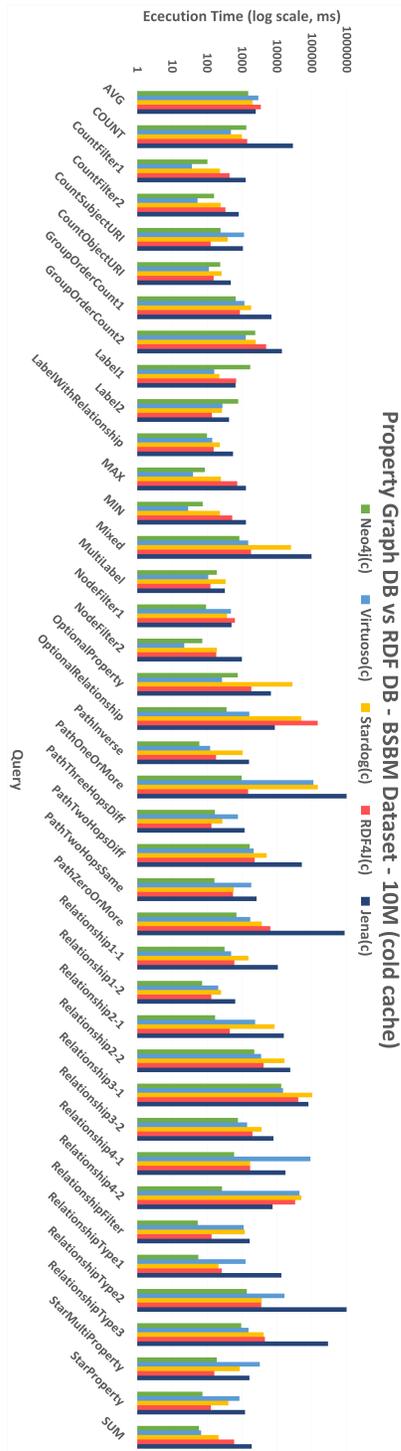} 
    \caption{Property graph database V.S. RDF database - BSBM Dataset\_10M (Cold Cache)}
    \label{fig:cold-cache}
\end{sidewaysfigure}

\begin{sidewaysfigure}
\centering
    \includegraphics[width=\linewidth]{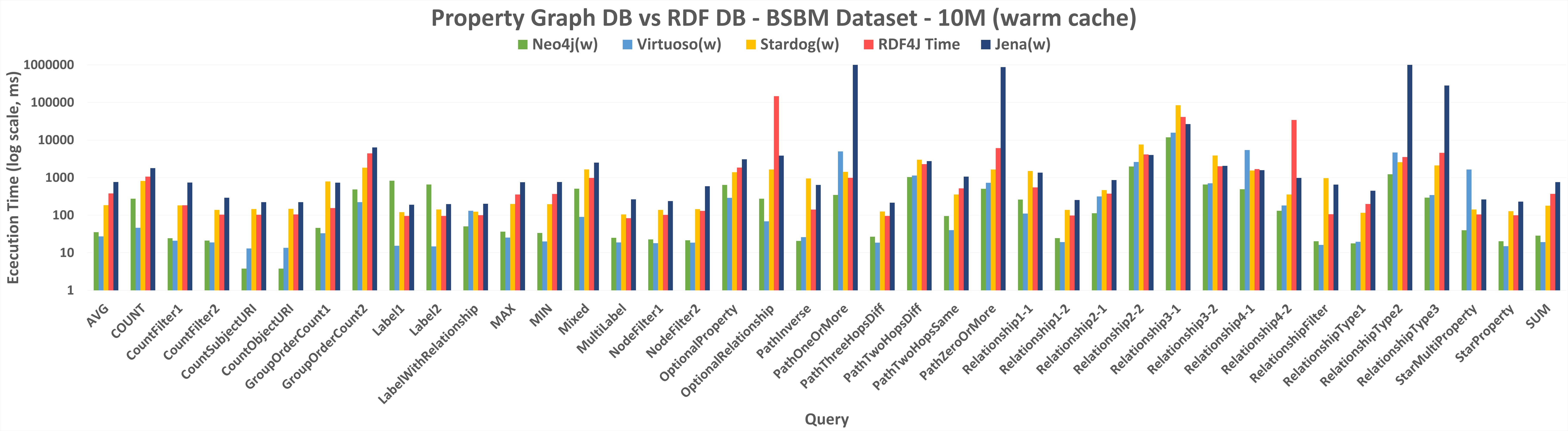} 
    \caption{Property graph database V.S. RDF database - BSBM Dataset\_10M (Cold Cache)}
    \label{fig:warm-cache}
\end{sidewaysfigure}

\end{document}